\documentclass[twocolumn,tighten]{aastex63}
\usepackage[caption=false]{subfig}
\usepackage{CJK} 

\newcommand{\logg}{$\log{g}$}

\submitjournal{ApJ}
\accepted{November 23, 2020}

\shorttitle{Ancient Very Metal-Poor Stars Associated With the Galactic Disk}

\graphicspath{{./}{figures/}}

\begin{document}
\begin{CJK*}{UTF8}{gbsn}

\title{Ancient Very Metal-Poor Stars Associated With the Galactic Disk in the H3 Survey}

\author[0000-0003-1144-7433]{Courtney Carter}
\affiliation{Grinnell College, 1115 8th Ave, Grinnell, IA 50112, USA}

\author[0000-0002-1590-8551]{Charlie Conroy}
\affiliation{Center for Astrophysics $|$ Harvard \& Smithsonian, 60 Garden Street, Cambridge, MA 02138, USA}

\author[0000-0002-5177-727X]{Dennis Zaritsky}
\affiliation{Steward Observatory, University of Arizona, 933 North Cherry Avenue, Tucson, AZ 85721-0065, USA}

\author[0000-0001-5082-9536]{Yuan-Sen Ting (丁源森)}
\altaffiliation{Hubble Fellow}
\affiliation{Institute for Advanced Study, Princeton, NJ 08540, USA}
\affiliation{Department of Astrophysical Sciences, Princeton University, Princeton, NJ 08544, USA}
\affiliation{Observatories of the Carnegie Institution of Washington, 813 Santa Barbara Street, Pasadena, CA 91101, USA}
\affiliation{Research School of Astronomy and Astrophysics, Mount Stromlo Observatory, Cotter Road, Weston Creek, ACT 2611, Canberra, Australia}

\author[0000-0002-7846-9787]{Ana Bonaca}
\affiliation{Center for Astrophysics $|$ Harvard \& Smithsonian, 60 Garden Street, Cambridge, MA 02138, USA}

\author[0000-0003-3997-5705]{Rohan P. Naidu}
\affiliation{Center for Astrophysics $|$ Harvard \& Smithsonian, 60 Garden Street, Cambridge, MA 02138, USA}

\author[0000-0002-9280-7594]{Benjamin D. Johnson}
\affiliation{Center for Astrophysics $|$ Harvard \& Smithsonian, 60 Garden Street, Cambridge, MA 02138, USA}

\author[0000-0002-1617-8917]{Phillip A. Cargile}
\affiliation{Center for Astrophysics $|$ Harvard \& Smithsonian, 60 Garden Street, Cambridge, MA 02138, USA}

\author[0000-0003-2352-3202]{Nelson Caldwell}
\affiliation{Center for Astrophysics $|$ Harvard \& Smithsonian, 60 Garden Street, Cambridge, MA 02138, USA}

\author[0000-0003-2573-9832]{Josh Speagle}
\affiliation{Center for Astrophysics $|$ Harvard \& Smithsonian, 60 Garden Street, Cambridge, MA 02138, USA}
\affiliation{University of Toronto, Department of Statistical Sciences, Toronto, M5S 3G3, Canada}
\affiliation{University of Toronto, David A. Dunlap Department of Astronomy \& Astrophysics, Toronto, M5S 3H4, Canada}
\affiliation{University of Toronto, Dunlap Institute for Astronomy \& Astrophysics, Toronto, M5S 3H4, Canada}

\author[0000-0002-6800-5778]{Jiwon Jesse Han}
\affiliation{Center for Astrophysics $|$ Harvard \& Smithsonian, 60 Garden Street, Cambridge, MA 02138, USA}

\begin{abstract}

Ancient, very metal-poor stars offer a window into the earliest epochs of galaxy formation and assembly. We combine data from the H3 Spectroscopic Survey and {\it Gaia} to measure metallicities, abundances of $\alpha$ elements, stellar ages, and orbital properties of a sample of 482 very metal-poor (VMP; [Fe/H]$<-2$) stars in order to constrain their origins. This sample is confined to $1\lesssim |Z| \lesssim3$ kpc from the Galactic plane. We find that $>70$\% of VMP stars near the disk are on prograde orbits and this fraction increases toward lower metallicities. This result unexpected if metal-poor stars are predominantly accreted from many small systems with no preferred orientation, as such a scenario would imply a mostly isotropic distribution. Furthermore, we find there is some evidence for higher fractions of prograde orbits amongst stars with lower [$\alpha$/Fe]. Isochrone-based ages for main sequence turn-off stars reveal that these VMP stars are uniformly old ($\approx12$ Gyr) irrespective of the $\alpha$ abundance and metallicity, suggesting that the metal-poor population was not born from the same well-mixed gas disk. We speculate that the VMP population has a heterogeneous origin, including both in-situ formation in the ancient disk and accretion from a satellite with the same direction of rotation as the ancient disk at early times. Our precisely measured ages for these VMP stars on prograde orbits show that the Galaxy has had a relatively quiescent merging history over most of cosmic time, and implies the angular momentum alignment of the Galaxy has been in place for at least 12 Gyr.

\end{abstract}


\section{Introduction} 
\label{sec:intro}

The properties of very metal-poor (VMP; [Fe/H] $<-2$) stars offer a unique window into the early formation and assembly history of the Galaxy. These stars are some of the oldest in the Milky-Way (MW); therefore, their chemistry provides insight into the earliest conditions of the Galaxy  \citep[e.g.,][]{Christlieb2002, Bland-Hawthorn2010, Frebel2015}. Simulations have shown \citep[e.g.,][]{Brook2007, Salvadori2010, ElBadry2018} that while the absolute number of metal-poor stars in MW-like galaxies is greatest in their central regions, the relative fraction is largest in the stellar halo. Simulations also predict that the majority of the ancient metal-poor stars in the Galaxy are accreted \citep{ElBadry2018}. These stars are generally predicted to be kinematically halo-like; if there is no preferred accretion orientation, these populations should be well-mixed, with no preference for prograde orbits. It is therefore expected that the ancient metal-poor stars in our Galaxy are likely to reside in the bulge and the stellar halo and have little to no net rotation. 

 If the distribution of VMP stars in the Galaxy has no preference for prograde orbits, then we should expect approximately half of these stars on prograde ($L_z<0$) orbits, and half on retrograde ($L_z>0$) orbits, but this is not the case. Recently, \citet{Sestito2020} found that a significant fraction ($\approx31$\%) of the VMP stars in their sample are kinematically associated with the disk (on prograde orbits). This discovery defies the conventional wisdom that the oldest populations of the Galaxy are either associated with the isotropic halo and/or accreted from a isotropic distribution of building blocks and suggests that (a) stars are preferentially formed or accreted in a similar configuration as the disk \citep[e.g,][]{Dubois2014} and (b) have not been heated sufficiently enough to completely erase their dynamical memory \citep[e.g,][]{Villalobos08}. A strong excess of VMP stars on prograde orbits likely means these stars were either born in-situ, or that ancient metal-poor stars found their way near the MW plane through processes such as gravitational torquing \citep{Codis_2012} and dynamical friction \citep{Read_2008}. Accretion from the cosmic web also tends to be aligned with the angular momentum of the star forming disk (preferentially prograde) \citep{Dubois2014}. 

Using data from the H3 Stellar Spectroscopic Survey in combination with \emph{Gaia}, we investigate the orbits of metal-poor stars with new insights from chemistry ($\alpha$-abundances) and ages. We confirm a large population of VMP stars on disk-like orbits near the Galactic midplane identified by \citet{Sestito2019,Sestito2020}. We find the population is uniformly old and more disk-like for the low-$\alpha$, VMP stars. Section \ref{sec:methods} describes the data and our sample selection. In Section \ref{sec:results}, we characterize the orbital properties of the VMP stars and provide isochrone ages for a subset of main sequence turn-off (MSTO) stars. In Section \ref{sec:discandcon} we discuss our results in the context of previous work and the formation history of the Galactic disk.

\section{Data} \label{sec:methods}

\begin{figure*}[!t]
    \centering
    \includegraphics[width=0.95\textwidth]{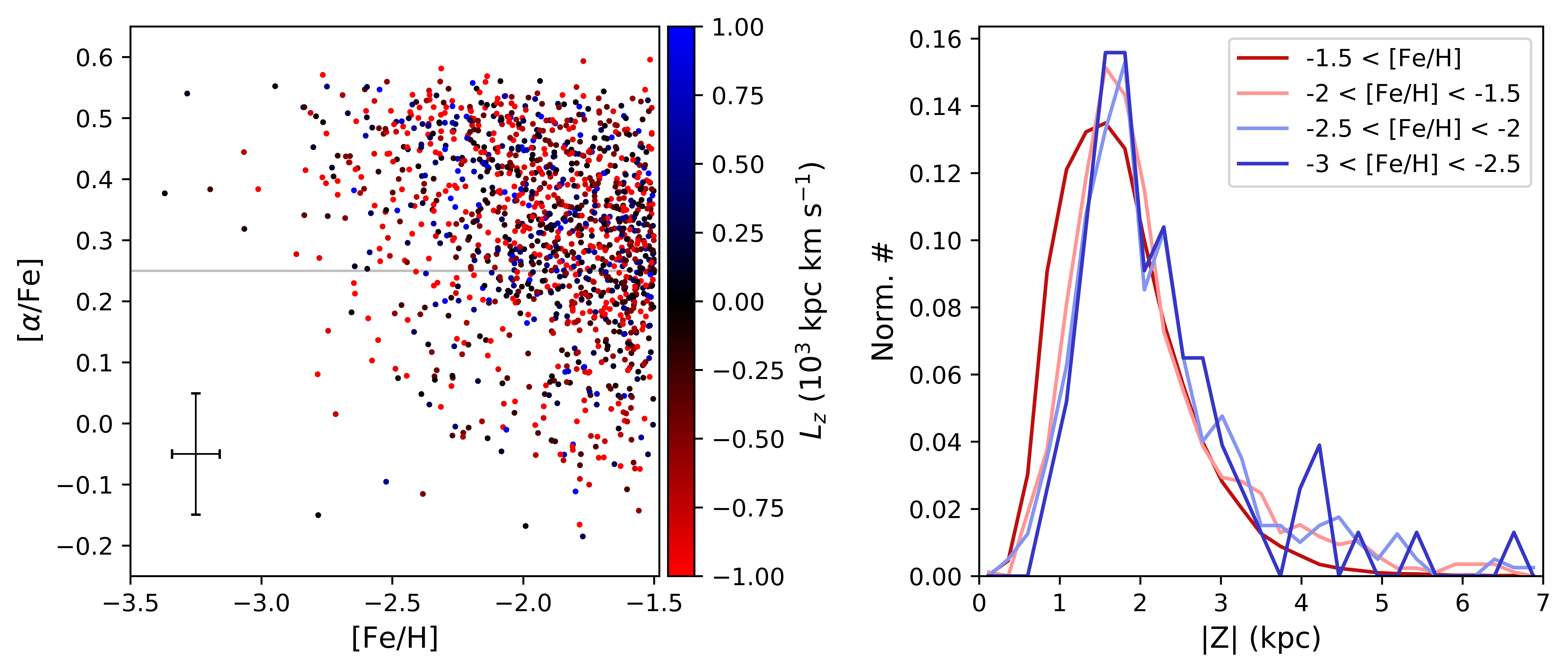}
    \caption{\emph{Left panel:} [$\alpha$/Fe] versus [Fe/H] for stars in the sample between $-3.5<$ [Fe/H] $<-1.5$, color-coded by $L_z$ (where prograde is $L_z<0$ and retrograde is $L_z>0$). A grey line is drawn at 0.25 [$\alpha$/Fe] for reference. \emph{Right panel:} Distribution of distances from the Galactic midplane for the metal-poor stars. Most of these metal-poor stars reside close to the disk (within $\approx$ 3 kpc). The lack of stars below 2 kpc is due to the dwarf selection function. There are 42101 stars with $-1.5 <$ [Fe/H], 852 with $-2 <$ [Fe/H] $< -1.5$, 399 with $-2.5 <$ [Fe/H] $< -2$, and 77 with $-3 <$ [Fe/H] $< -2.5.$ The remaining 6 VMP stars not included in the figure have [Fe/H] $< -3.$}
    \label{fig:plot1}
\end{figure*}

\begin{figure*}[!htbp]
    \centering
    {{\includegraphics[width=0.95\textwidth]{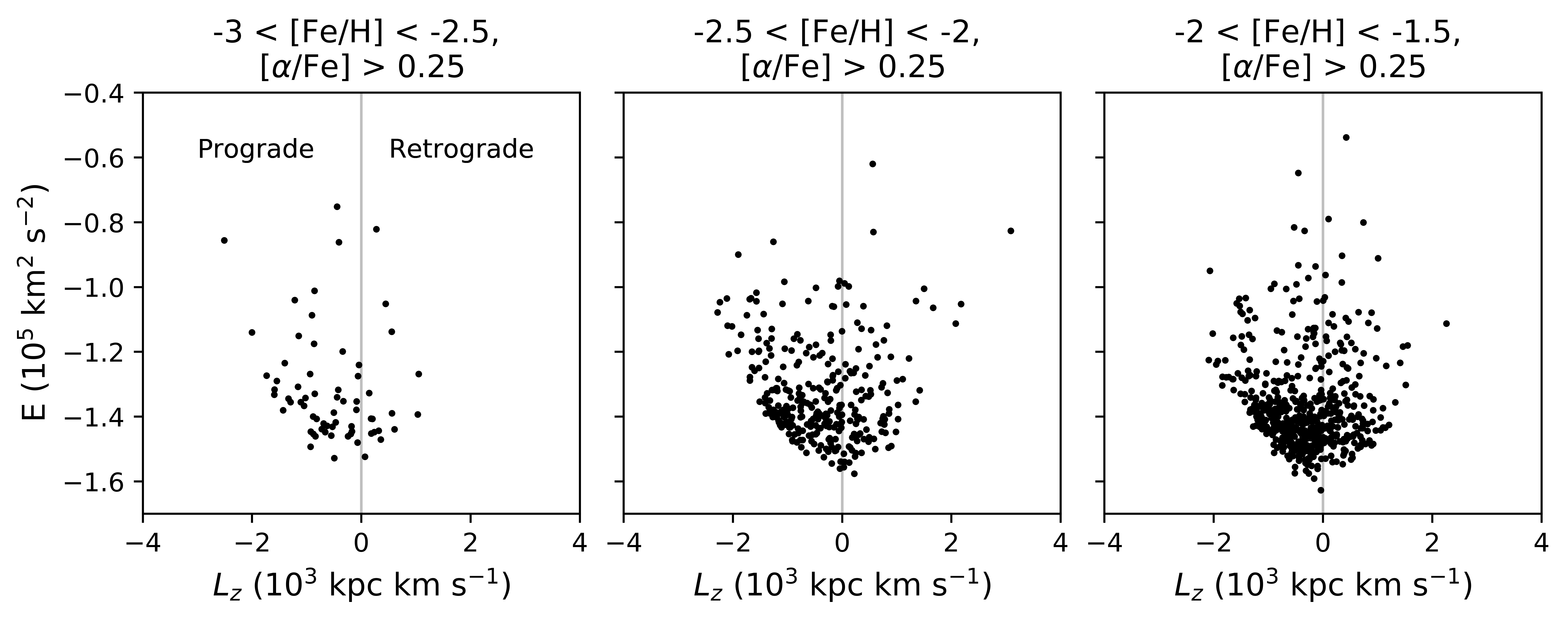} }}
    \qquad
   {{\includegraphics[width=0.95\textwidth]{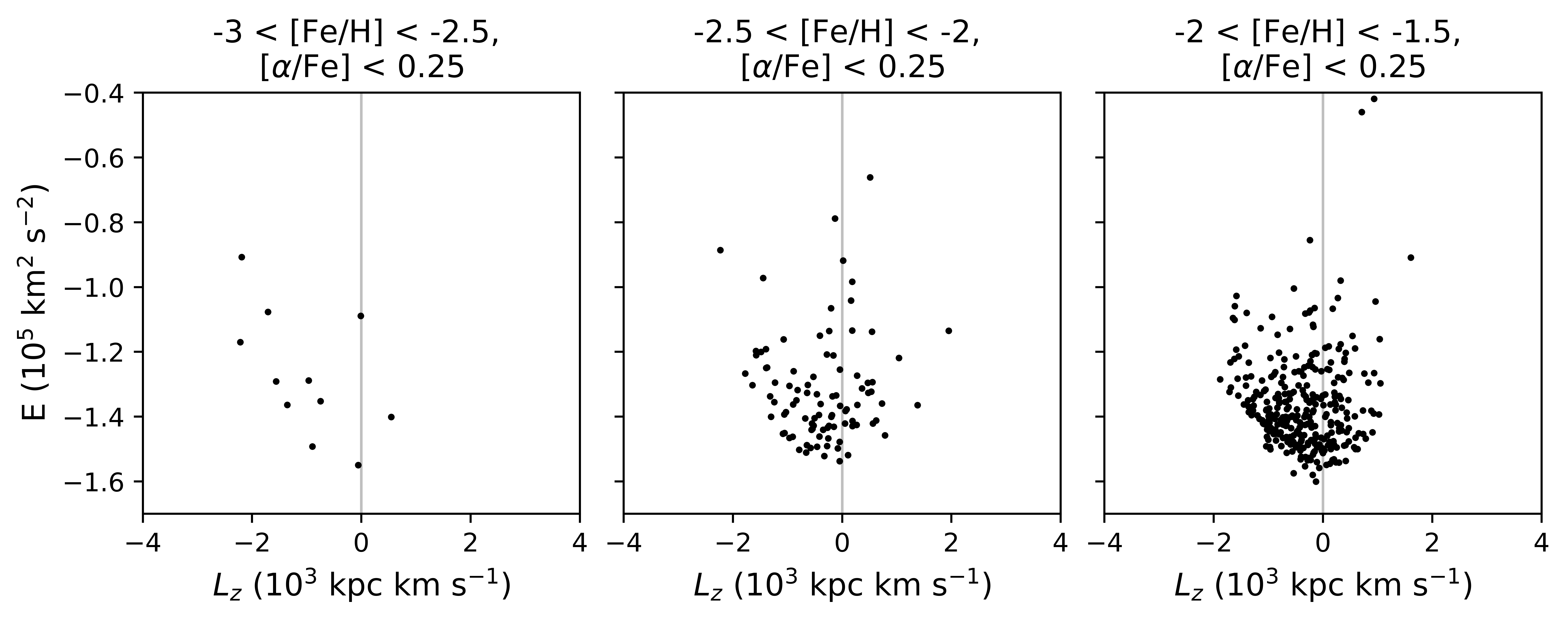}}}
    \caption{\emph{Top panels:} $E - L_z$ space for high-$\alpha$ ([$\alpha$/Fe] $>$ 0.25) stars in  0.5 dex bins from $-3$ to $-1.5$ in [Fe/H]. An isotropic distribution of metal-poor halo stars should have no net rotation. Unexpectedly, the most metal-poor stars have the highest fraction of stars on prograde orbits. \emph{Bottom panels:} Similarly to the top panel, $E - L_z$ space for low-$\alpha$ ([$\alpha$/Fe] $<$ 0.25) stars in 0.5 dex bins. Interestingly, an even greater fraction of the VMP stars are on prograde orbits for the low-$\alpha$ population. At the lowest metallicity bin ($-3 <$ [Fe/H] $< -2.5$) ten out of twelve stars are prograde-moving.}
    \label{fig:plot2}
\end{figure*}

The H3 Survey \citep{Conroy2019} is an ongoing high-resolution spectroscopic survey optimized for observing the stellar halo.  Specifically, the selection function for the main sample consists of a magnitude selection: $15<r<18$ and a parallax selection of $\varpi - 2\sigma_\varpi<0.5$ for the first half of the survey and $\varpi<0.4$ mas for the later half of the survey. Survey fields are located at high Galactic latitudes: $\vert b \vert>30^{\circ}$ and Dec. $>-20^{\circ}$. Spectra are obtained via the high-resolution fiber-fed Hectochelle spectrograph on the MMT. Hectochelle is configured for H3 to deliver $R\approx32,000$ spectra over the wavelength range $5150-5300$\AA. Strong Fe lines and the Mg I triplet dominate in the wavelength range of Hectochelle. H3 has collected spectra of 130,000 stars to-date with a goal of reaching $\approx200,000$.

Stellar parameters including [Fe/H] and [$\alpha$/Fe], distance, radial velocity, mass, and age are measured with \texttt{MINESweeper} \citep{Cargile2019}.  \texttt{MINESweeper} determines physical stellar parameters by using \texttt{MIST} \citep{Choi2016} stellar isochrones to determine the evolutionary state, mass, and initial bulk composition of stars\footnote{The initial, bulk metallicity of a star can differ from its present surface abundance due to mixing and diffusion.  Both of these processes are included in the \texttt{MIST} isochrones.  The isochrone surface abundance is used to determine the appropriate corresponding spectral model.} which are then ``matched'' to both model spectra and photometry using a custom grid of custom stellar spectral models. \texttt{MINESweeper} uses this grid of models to fit both the Hectochelle spectrum and the broadband photometric spectral energy distribution (SED) using a Bayesian framework. {\it Gaia} parallaxes and a Galactic density and age model are included as priors. See \citet{Conroy2019} and \citet{Cargile2019} for more details. The typical precision on derived parameters are $\lesssim 1$ km s$^{-1}$ for radial velocities, $\lesssim0.1$ dex for [Fe/H] and [$\alpha$/Fe], and $\lesssim$ 10$\%$ for spectrophotometric distances.  We note that while [$\alpha$/Fe] is the fitted parameter, we are most sensitive to [Mg/Fe] as the Hectochelle spectrum contains very strong Mg lines.  

With 3D positions and 3D velocities we compute a variety of derived quantities including Galactocentric positions and velocities and angular momenta. Orbit-based quantities are computed with an assumed model for the Galactic potential. For this purpose we use the \texttt{gala v1.1} package \citep{Price_Software_2017, Price-Whelan2017} with its default \texttt{MilkyWayPotential}. Orbits are computed using the explicit integration scheme by \citet{Dormand1978} with a total integration time of 25 Gyr in time-steps of 1 Myr. The H3 sample contains halo stars on long period orbits and this integration time is more than sufficient to ensure well-converged orbital parameters. We report total orbital energy, $E_{tot}$, in units of $10^5$ $\textrm{km}^2$ $\textrm{s}^{-2}$ and angular momenta ($L_x, L_y, L_z$) in units of $10^3$ kpc km $\textrm{s}^{-1}$. We use the Galactocentric frame of reference provided by \texttt{Astropy v4.0}, in which prograde (aligned with the current angular momentum vector of the Galaxy) orbits have $L_z<0$ and retrograde (anti-aligned) orbits $L_z>0$. The mean uncertainties for $\sigma_{E_{\rm tot}} \approx 0.007 \times 10^5$ $\textrm{km}^2$ $\textrm{s}^{-2}$ and for $\sigma_{L_z} \approx 0.037 \times 10^3$ kpc km $\textrm{s}^{-1}$. See \citet{Naidu2020} for a more in-depth explanation of how phase space quantities are computed.

The sample analyzed in this paper is selected to have SNR$>7$ and no data quality flags. The latter excludes from our sample a small number of spectra with extraction problems, poorly-fit stellar parameters, and/or or SNR$<1$. The SNR selection is higher than used in some previous H3 work because here we focus on the most metal-poor stars in the survey; a good measurement of metallicity and other stellar parameters requires a higher SNR threshold. The main results (summarized in Figures $2-4$) do not change if we adopt a SNR threshold of 5 or 10. We also select a well defined region of dwarf stars (log $g>3.5$), limiting the sample to stars within a few kpc of the Galactic plane. We removed stars with large uncertainties in $L_z$: $\sigma_{L_z}>0.2\times10^3$ kpc km s$^{-1}$. This cut removes $\approx3\%$ of the stars from the final overall sample, which contains 43,435 stars. Finally, inspection of the subset of 492 VMP stars by eye revealed 10 with evidence of problems with either the data or the fitted solution were removed from the sample. What remains is our final sample containing 482 VMP stars.

\section{Results} \label{sec:results}

\begin{figure*}[t!]
    \centering
     \includegraphics[width=0.47\textwidth]{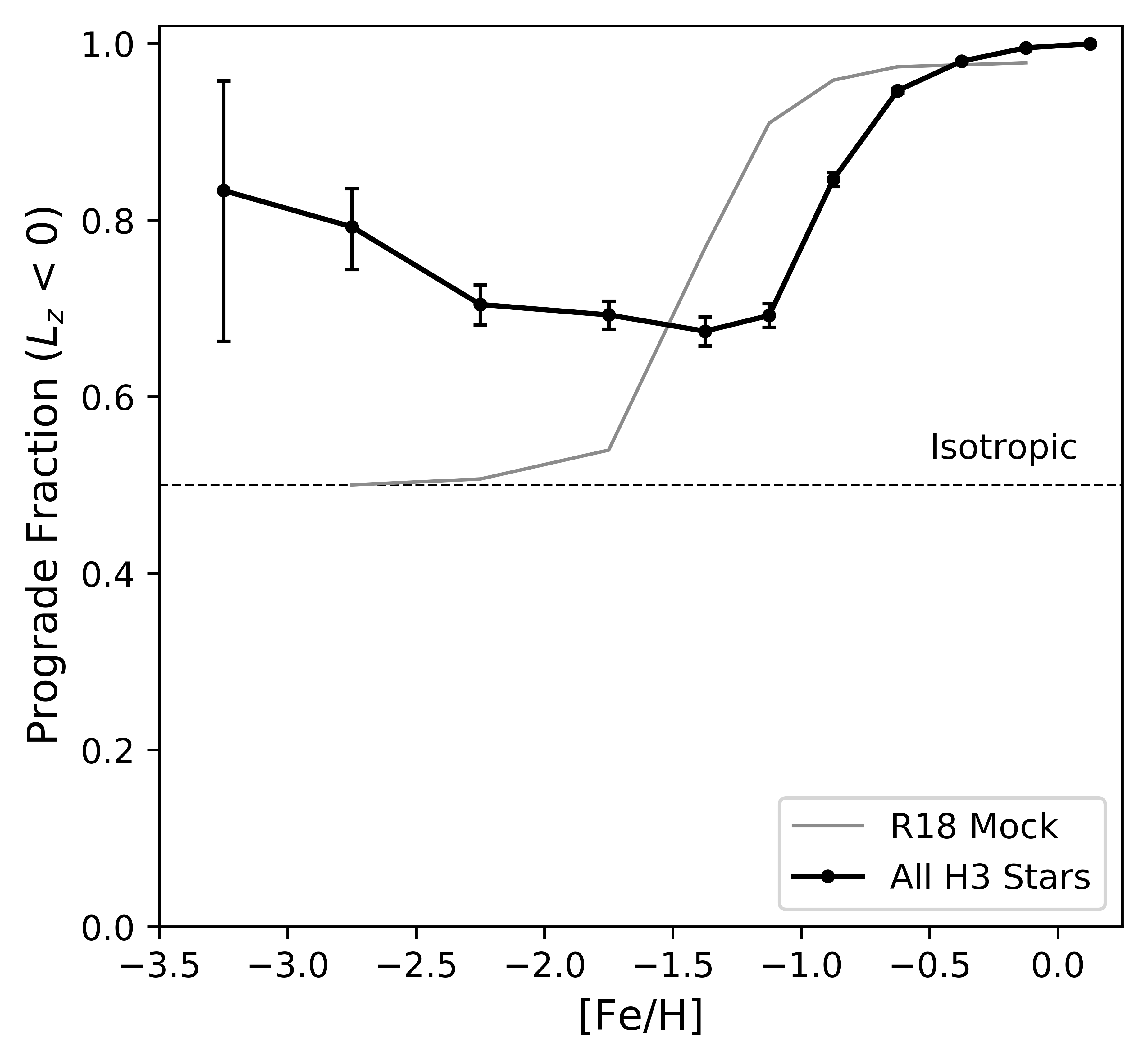}
    \includegraphics[width=0.47\textwidth]{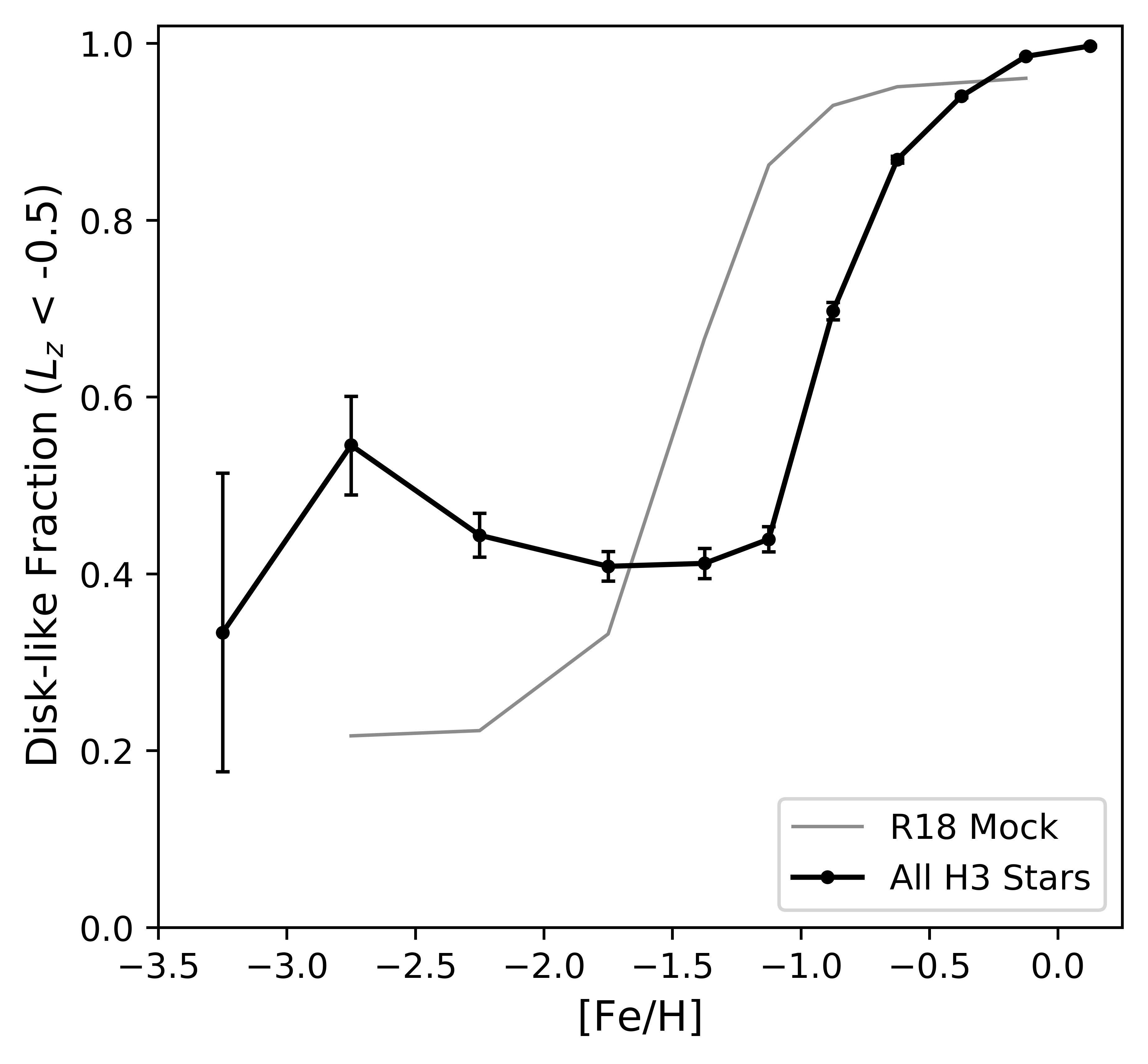}
    \includegraphics[width=0.47\textwidth]{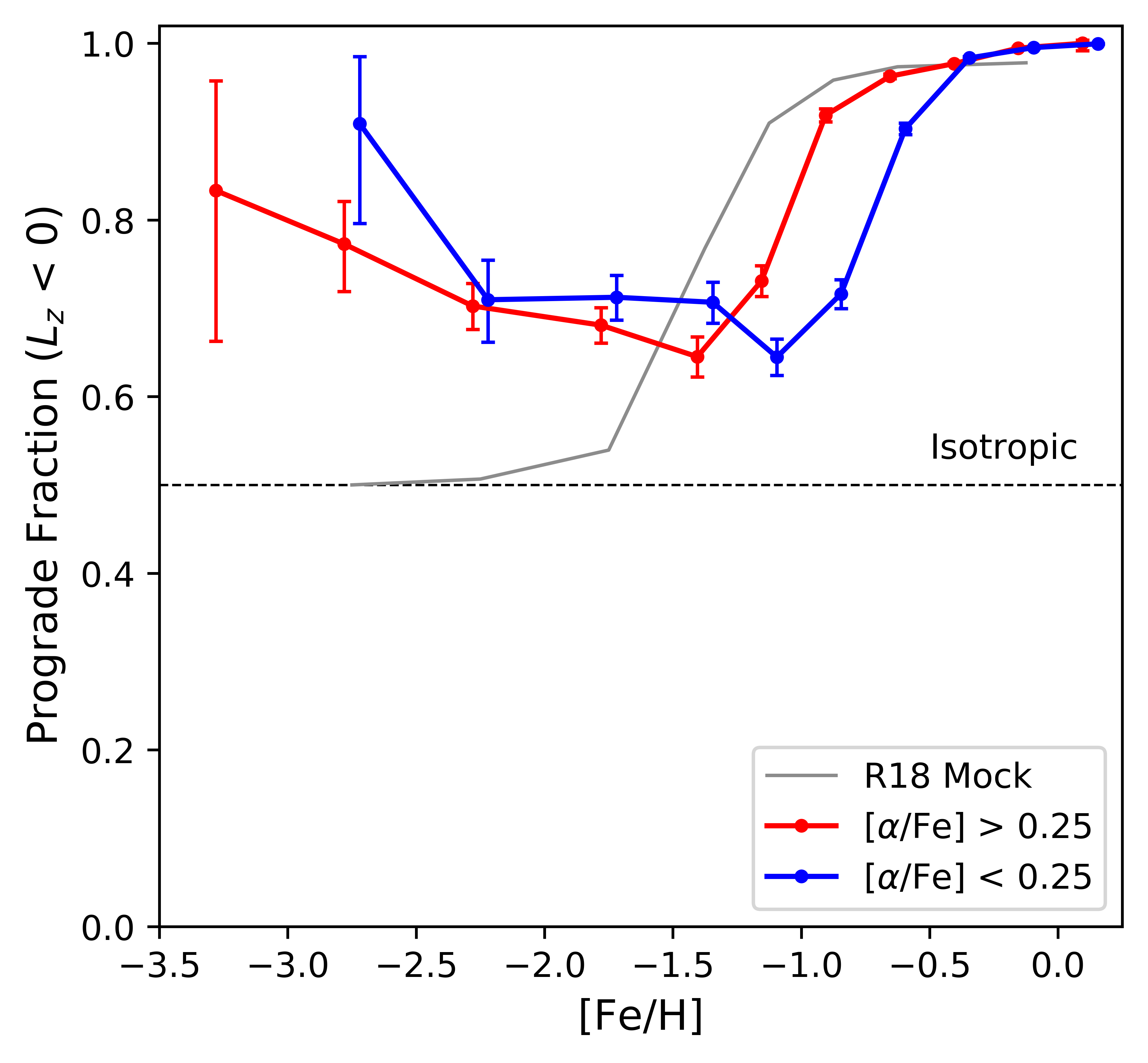}
    \includegraphics[width=0.47\textwidth]{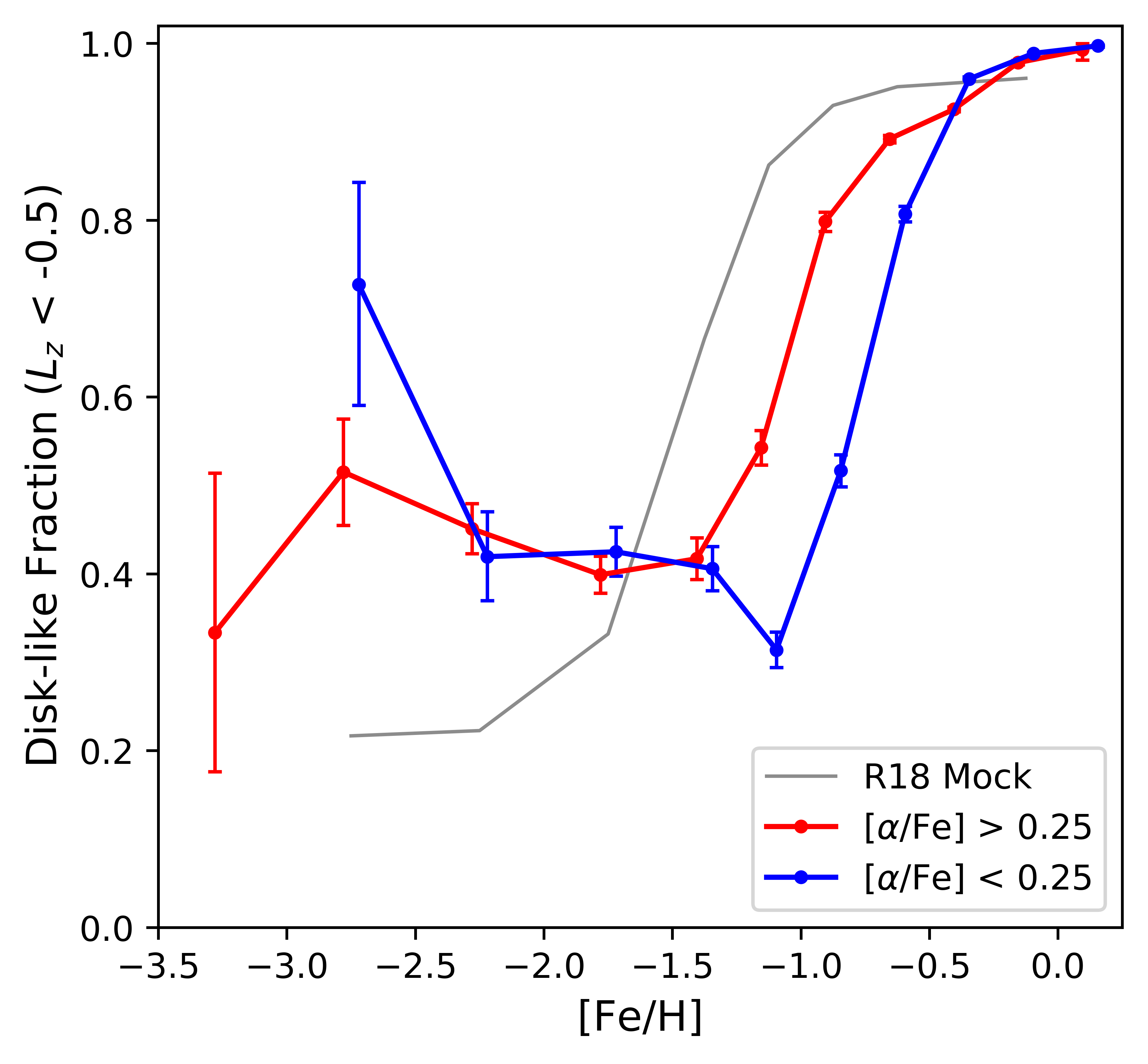}
    \caption{\emph{Left panels:} Fraction of stars on prograde orbits as a function of metallicity. Top panel includes all sample stars. Results in the bottom panel are shown separately for high and low$-\alpha$ stars in red and blue, respectively. As expected, the stars become more disk-like with prograde rotation at higher metallicities. Surprisingly, the prograde fraction increases again for the VMP stars. Gray lines feature stars from the \citet[][R18]{Rybizki2018} mock MW catalogue (characterized by a smooth disk and halo) for comparison. In R18 the VMP stars are dominated by halo stars on isotropic orbits, and hence the prograde fraction converges to $\approx0.5$. \emph{Right panels:} Similar to the left panels, but now for the disk-like fraction of stars as a function of metallicity. The disk-like fraction of the low-$\alpha$ stars mirrors the bottom left panel, while the high-$\alpha$ stars are consistent with a flat ratio at [Fe/H] $<-1.5$.}
    \label{fig:plot3}
\end{figure*}

We begin with an overview of the sample in Figure \ref{fig:plot1}. The left panel shows the distribution of stars in [$\alpha$/Fe] versus [Fe/H] for [Fe/H]$<-1.5$. Inspection of this panel shows a fair number of stars above and below [$\alpha$/Fe] of 0.25 which motivates our working definition of high and low-$\alpha$ for the VMP stars. In general, the more metal-poor stars have higher [$\alpha$/Fe] ratios, as expected from previous work \citep[e.g.,][]{Carretta2000}. and chemical evolution models \citep[e.g.,][]{Kobayashi_2006}. Curiously, there is an additional interesting population of VMP, low-$\alpha$ stars. It is also important to note there are no extremely metal-poor ([Fe/H] $< -2.9$) stars with low-$\alpha$ in our sample. Average error bars are plotted along the bottom.

The right panel of Figure \ref{fig:plot1} shows the distribution of absolute distances, $|Z|$, from the Galactic plane, for stars of different metallicities. The overall distribution is driven largely by (a) the selection function (specifically the parallax cut), and (b) the dwarf selection/cut, such that stars in the sample are confined to $1\lesssim |Z| \lesssim3$ kpc from the Galactic plane. There are no strong differences in $|Z|$ as a function of metallicity for the most metal-poor stars in our sample ([Fe/H]$<-1.5$).

In Figure \ref{fig:plot2} we show the distribution of stars in orbital energy ($E$) and the $z-$component of angular momentum ($L_z$) space.  These two quantities should be approximately conserved in a static axisymmetric potential; $E-L_z$ space has therefore become a popular projection in which to view stellar populations in the Galaxy. Each column contains a narrow range in metallicities from $-3.5$ to $-1.5$ in 0.5 dex increments. The top panel shows high-$\alpha$ stars ([$\alpha$/Fe]$> 0.25$) and the bottom panel shows low-$\alpha$ stars ([$\alpha$/Fe]$<0.25$). The more metal-rich panels show evidence of the \emph{Gaia}-Sausage-Enceladus (GSE) accreted dwarf galaxy \citep{Belokurov2018, Helmi2018}, composed of a population of stars at $L_z\approx0$. An unexpected result is the large fraction of stars on preferentially prograde orbits ($L_z<0$) at the lowest metallicities \citep[a result first reported in][]{Sestito2019, Sestito2020}. This is most dramatic in the lower left panel, which shows a mostly prograde-moving population of ten out of twelve low-$\alpha$, metal-poor stars. One unbound star ($E>0$) is not shown in the last panel.

The key result of this work is shown in Figure \ref{fig:plot3}. In the left panel we show the fraction of stars on prograde orbits ($L_z<0$) as a function of [Fe/H] for both the high and low-$\alpha$ stars. We also show the predictions from the \citet[][R18]{Rybizki2018} mock GDR2 catalog, matched to the dwarf selection function in this study. R18 employs smooth models for the stellar thin and thick disks and halo, making it a useful aid for comparison with the data in our sample.\footnote{The metallicities of halo stars in R18 were increased by 0.55 dex to agree with the average metallicity of the Galactic stellar halo \citep{Conroy2019b}.} The right panel shows the fraction of stars on disk-like orbits ($L_z<-0.5\times10^3$ kpc km s$^{-1}$, which corresponds to $e\lesssim0.7$ in our sample).\footnote{The uncertainties shown in Figure \ref{fig:plot3} were calculated assuming a binomial distribution with a uniform prior, where the posterior is defined by $P(p | k, N) \propto P(k | p, N) P(p)$ and $p$ is the true underlying fraction of prograde stars, $k$ is the observed number of prograde stars, and $N$ is the total number of stars in the bin.  The uncertainties measure the likelihood of $k$ given $N$, with a probability of $p$. This means that a sample in which all stars have $L_z<0$ will not necessarily result in a prograde fraction of one; the uncertainties measure the likelihood of observing this fraction of prograde stars, not the likelihood of the stars being prograde.}

The R18 model forms a useful baseline for what we might expect to observe in Figure \ref{fig:plot3}. At high metallicities the prograde fraction rises to 1.0, reflecting the dominant contribution of disk stars at high metallicities. At lower metallicities the fraction drops to 0.5, reflecting the dominant contribution from halo stars with no net rotation.

\begin{figure*}[t!]
    \centering
    \includegraphics[width=0.95\textwidth]{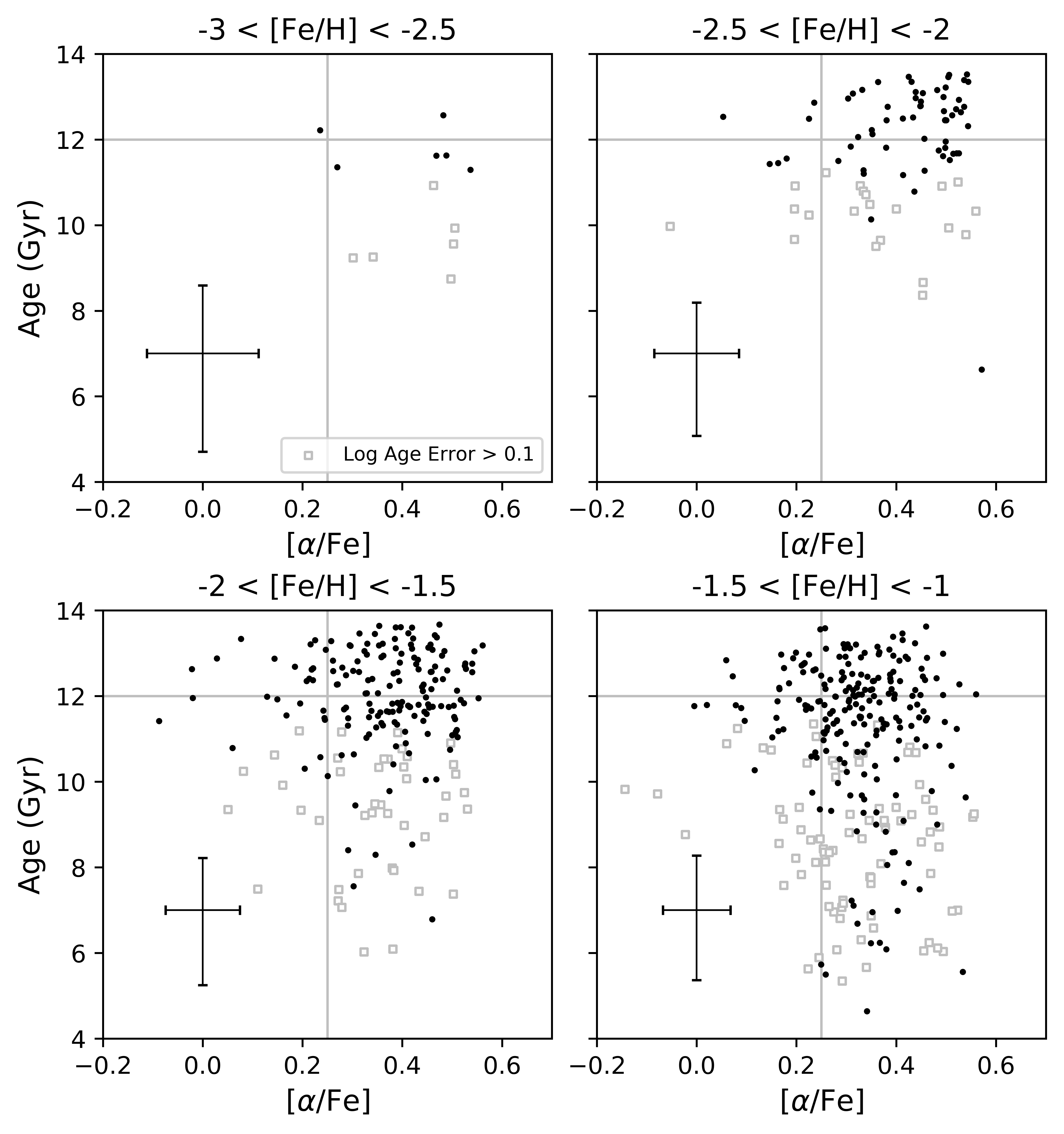}
    \caption{Age versus [$\alpha$/Fe] for stars in four metallicity bins. Stars in this plot are restricted to those on or near the main sequence turnoff ($3.5<$\logg$<4.3$) where more robust isochrone ages can be attained. Grey square points highlight stars with a log age error greater than 0.1. Mean error bars are shown in the bottom left-hand corner of each plot. Stars with log age error $>$ 0.1 are not included in the average error bars. A grey line at 12 Gyr and at 0.25 [$\alpha$/Fe] is shown in each panel for reference. Notably, when excluding stars with large age uncertainties, there are no prominent trends in age with [$\alpha$/Fe] for the most metal-poor stars. This is in contrast to the behavior seen in stars at higher metallicities \citep[e.g.,][]{Haywood2013, Bonaca2020}. For comparison, the fraction of all stars with ages $< 10$ Gyr jumps from about $16\%$ to $34\%$ from the bottom left and right panels at $-2 <$ [Fe/H] $< -1.5$ and $-1.5 <$ [Fe/H] $< -1$ respectively. Note, this plot differs from those shown in other work due to the H3 selection function, which misses younger stars near the Galactic plane.}
    \label{fig:plot4}
\end{figure*}

As expected, the data shows a high fraction of stars on prograde orbits at [Fe/H]$>-0.75$ and a rapid drop in prograde fraction over the range $-1.5<$[Fe/H]$<-0.75$ because metal-poor stars are preferentially associated with the halo. The downturn at -1.0 [Fe/H] occurs due to GSE stars, which have a metallicity distribution that peaks at [Fe/H] $\approx-1.2$. These stars generally have low-[$\alpha$/Fe] \citep{Conroy2019b, Naidu2020}, which explains why the low$-\alpha$ prograde fraction drops at higher metallicity than the high$-\alpha$ population.

Surprisingly, the fraction of stars on prograde orbits {\it increases} with decreasing metallicity below [Fe/H] $\approx-1.5$, reaching values of $\gtrsim0.8$ at the lowest metallicities in the sample. In addition, the prograde fraction of the low$-\alpha$ stars is somewhat higher than the high$-\alpha$ stars, especially at the lowest metallicities. The right panels of Figure \ref{fig:plot3} show the fraction of stars on disk-like orbits, and the comparison to R18 further highlights that such a large fraction of stars on prograde, disk-like orbits is unexpected if the lowest-metallicity stars were associated with an isotropic stellar halo.

We experimented with several definitions of high and low$-\alpha$ (including a diagonal separation in the [Fe/H] vs. [$\alpha$/Fe] plane). The rising fraction of stars on preferentially prograde orbits at the lowest metallicities, seen in Figure \ref{fig:plot3}, is independent of these selections. However, quantitative differences between the high and low$-\alpha$ populations are sensitive to this choice.

Ages offer an important clue to the origin of these metal-poor stars on prograde orbits. In Figure \ref{fig:plot4} age is plotted against [$\alpha$/Fe], where each panel focuses on a narrow range in metallicities. In this figure we restrict the sample to stars near the main sequence turn-off (MSTO) or sub-giant branch by requiring $3.5<$ \logg $<4.3$. This range of \logg\ produces isochrone ages with the smallest uncertainties. We computed these ages without the influence of a Galactic age prior \citep[See also][]{Bonaca2020}. Mean error bars have been plotted in the bottom left-hand corner of each panel. For [Fe/H]$>-1$, there is a well-established relation between age and [$\alpha$/Fe] such that higher$-\alpha$ corresponds to older ages \citep[e.g.,][]{Haywood2013, Bonaca2020}. With increasing metallicity ranges, we see a larger and larger spread in ages. In the bottom right of Figure 4 ($-1.5 <$ [Fe/H] $< -1$), this is evident by the appearance of high$-\alpha$ stars at ages as young as $\approx 5$ Gyr. No such trend is apparent at the lowest metallicities shown in Figure \ref{fig:plot4}. This suggests that the low and high$-\alpha$ populations at low-metallicity formed contemporaneously, yet they must have different origins.

\vspace{1cm}

\section{Discussion} \label{sec:discandcon}

It is remarkable that such a large fraction of metal-poor stars are on prograde (disk-like) orbits, especially given that the H3 selection function preferentially {\it avoids} the Galactic midplane (see Figure \ref{fig:plot1} right panel). We find that $>70$\% of VMP stars near the disk are on prograde orbits and this fraction increases toward lower metallicities. The prograde fraction may be even higher if stars were observed with smaller $|Z|$.  The survey selection of $|b|>30^\circ$ and $r>2$ kpc also ensures we avoid the bulge, where simulations \citep[e.g.,][]{ElBadry2018}) predict that a large number of VMP stars should reside. For these reasons our observed fraction of VMP stars on prograde orbits is likely not representative of the entire population of VMP stars in the Galaxy.

Recent work by \citet{Sestito2019, Sestito2020} has demonstrated that a significant fraction of metal-poor stars are on prograde orbits confined to the disk.  These authors considered several possible scenarios for the origin of such metal-poor prograde stars, including both in-situ (formed within the ancient disk of the Galaxy) and ex-situ (accreted through minor mergers or along with the building blocks of the proto-Galaxy) channels.  In this work we use data from the H3 Survey to investigate the orbital properties of metal-poor stars within 3 kpc, confirming the conclusion of Sestito et al. that a large majority of VMP stars are on prograde-orbits (these authors also show that many prograde stars have orbits confined to the plane --- the H3 selection function avoids such stars). In addition, we were able to study this effect as a function of $\alpha$ abundances, finding evidence for a slightly higher fraction of prograde-orbits amongst the low$-\alpha$ population. A subset of our sample lies near the MSTO, which allows for precise and accurate ages. These stars are uniformly old ($\approx12$ Gyr), independent of both [Fe/H] {\it and} [$\alpha$/Fe].  

Our preferred scenario links these metal-poor stars to the very early formation and assembly of the Galaxy.  We observe uniformly old ages and a wide range in [$\alpha$/Fe] at fixed [Fe/H].  In a one-zone model of chemical evolution, these three variables are strongly linked to one another such that metallicity increases and [$\alpha$/Fe] decreases with decreasing age.  We do not observe this in our data, which suggests that these stars were not all born from the same (well-mixed) system.  We speculate that the high$-\alpha$ stars were born in the Galaxy, while the low$-\alpha$ stars are the result of slower-paced chemical evolution, which suggests they were born in a shallower potential well and were accreted with angular momentum that was generally aligned with the Galactic disk. The VMP, low$-\alpha$ stars remain puzzling. Previous work \citep{Nagasawa2018} explores the origins of chemically similar stars and suggests VMP, low$-\alpha$ stars could also be explained by early dominance of Type Ia supernovae (SN) or yields of a single rare nucleosynthetic event (e.g., a pair-instability SN) during the early formation of the Galaxy.

An alternate scenario is the late-time accretion of dwarf galaxies on prograde orbits \citep[e.g.,][]{Abadi03, Scannapieco2011}.  For such a scenario to be viable, the dwarf(s) must have very low average metallicity (e.g., [Fe/H]$\lesssim-2$), which corresponds to dwarf galaxy stellar masses of $<10^6\, M_\odot$ \citep{Kirby2013}.  Observations of local dwarf galaxies have found that the relation between [Fe/H] and [$\alpha$/Fe] is relatively narrow within a given galaxy \citep[See recent review from][]{Simon2019}. It is therefore unlikely that a single dwarf galaxy could produce the observed wide range in [$\alpha$/Fe] at low metallicities shown in this work. Multiple metal-poor dwarfs accreted on prograde-orbits would be required to satisfy the existing constraints. Whether this is likely can be tested by comparison to simulations.

The origins of VMP stars in similar galaxies has been recently explored using the NIHAO-UHD simulations \citep{Sestito_2020_Sims}. This work confirms the possibility of an early-accretion scenario while the proto-galaxy was undergoing a barrage of mergers and funnelling VMP stars to the innermost halo of the host galaxy. The simulations also find VMP stars originating from late-time mergers, appealing to a preference for the accretion of stars with prograde orbits at late-times. There was no evidence in support of VMP stars forming within an ancient disk. However, neither of these scenarios specifically address the observed differences in $\alpha$-abundance, leaving the interpretation of the high and low-$\alpha$ differences in disk-like, VMP populations as an open question. The simulations show a large majority of VMP stars fall between ages of 12-13 Gyr, in good agreement with our results.

Isochrone-based ages and chemistry from the H3 Survey confirm the existence of ancient, VMP stars near the disk. Since these stars have not been heated enough to significantly alter their preferentially prograde orbits, the angular momentum alignment of the Galaxy has likely been in place for at least 12 Gyr; this is possible only if the Galaxy has had a relatively quiescent merging history over most of cosmic time.  Indeed, {\it Gaia} has revealed that the last major merger of the Galaxy was likely GSE, which probably had a mass ratio of 1:4 or less \citep{Belokurov2018, Helmi2018, Naidu2020}, and such a merger is not expected to completely disrupt the disk \citep{Villalobos08}.

Future work would benefit from targeted observations of these stars to provide both more precise values for [Fe/H] and [$\alpha$/Fe] as well as more detailed abundance patterns, especially for comparison with theoretical SN yield models. In particular, neutron capture elements could help to distinguish various formation scenarios \citep[e.g.,][]{Alexander2019}. Detailed SN enrichment models, such as the recent work of \citet{Muley2020}, will play a role in interpreting the observed spread in [$\alpha$/Fe] for stars at the lowest metallicities. Finally, a large suite of simulations focused on the early formation and assembly of the Galaxy would provide valuable insight into the likely complex interplay between in-situ star formation and rapid accretion and merging that supports the current cosmological model.

\acknowledgments
\label{sec:acknowledgments}

We thank the Hectochelle operators Chun Ly, ShiAnne Kattner, Perry Berlind, and Mike Calkins, and the CfA and U. Arizona TACs for their continued support of the H3 Survey. Observations reported here were obtained at the MMT Observatory, a joint facility of the Smithsonian Institution and the University of Arizona. Y.S.T. is grateful to be supported by the NASA Hubble Fellowship grant HST-HF2-51425.001 awarded by the Space Telescope Science. This work has made use of data from the European Space Agency (ESA) mission
{\it Gaia} (\url{https://www.cosmos.esa.int/gaia}), processed by the {\it Gaia}
Data Processing and Analysis Consortium (DPAC,
\url{https://www.cosmos.esa.int/web/gaia/dpac/consortium}). Funding for the DPAC
has been provided by national institutions, in particular the institutions
participating in the {\it Gaia} Multilateral Agreement.

\vspace{5mm}
\facilities{MMT (Hectochelle), \emph{Gaia}}

\software{\texttt{IPython \citep{Perez2007}, matplotlib \citep{Hunter2007}, numpy \citep{van2011numpy}, Astropy \citep{astropy:2018}, SciPi \citep{2020SciPy-NMeth}, MINESweeper \citep{Cargile2019}, gala v1.1 \citep{Price_Software_2017,Price-Whelan2017}}}

\bibliographystyle{aasjournal}

\end{CJK*}
\end{document}